\documentclass[aps,prb,floatfix,twocolumn,showpacs,10pt,longbibliography]{revtex4-2}
\usepackage{amsmath}
\usepackage{braket} 
\usepackage{url}
\usepackage[ruled,vlined]{algorithm2e}
\usepackage{algorithmic}
\usepackage{graphicx}
\usepackage{dcolumn}
\usepackage{bm}
\usepackage{amssymb}
\usepackage{rotating}
\usepackage[abs]{overpic}
\usepackage{xcolor}
\usepackage{booktabs}
\usepackage{tikz}
\usepackage{tabularx}
\usepackage{ragged2e}
\usepackage{hyperref}
\hypersetup{colorlinks = true, citebordercolor={blue}, linkcolor={blue}, citecolor={blue}, urlcolor={blue}}

\DeclareUnicodeCharacter{2212}{-}
\DeclareUnicodeCharacter{2009}{-}

\begin{document}


\title{Witnessing Entanglement in Mixed-Particle Quantum Systems}
\author{Irma Avdic and David A. Mazziotti}

\email{damazz@uchicago.edu}

\affiliation{Department of Chemistry and The James Franck Institute, The University of Chicago, Chicago, IL 60637 USA}

\date{Submitted December 19, 2025}

\begin{abstract}
We introduce an entanglement witness that identifies off-diagonal long-range order (ODLRO)---a distinctive form of entanglement---in systems containing both fermionic and bosonic particles. By analyzing the particle–hole reduced density matrices of each subsystem, the approach detects ODLRO independently in both fermionic and bosonic sectors and identifies when long-range order develops across the entire mixed-particle system. The witness also quantifies the magnitude of ODLRO within each particle type, revealing how fermionic and bosonic correlations combine to form the total entanglement of the system, including a bosonic condensation of particle–hole pairs driven by many-body correlations rather than particle statistics. Using the Lipkin–Meshkov–Glick spin model, we show how the transition from ODLRO localized to one particle type to ODLRO shared by both particle types captures the onset of collective entanglement in a mixed-particle environment, providing new insight into systems where fermionic and bosonic correlations coexist.
\end{abstract}

\maketitle

{\em Introduction---} Entanglement, often described as the most non-classical manifestation of quantum mechanics~\cite{Schrodinger1935,Wootters1998,Bell2004,Horodecki2009}, plays a major role as a resource enabling various quantum technologies, ranging from quantum computing and cryptography~\cite{Nielsen_Chuang_2011,Jozsa2003,Yin2020,Naik2000} to metrology and sensing~\cite{Degen2017,Avdic2023,Chin2025,Prabhu2025,Giovannetti2011,DemkowiczDobrzaski2014,Toth2012}. It not only persists in large many-body systems and at finite temperatures~\cite{Amico2008} but can also be revealed through macroscopic observables such as heat capacity~\cite{Singh2013} and magnetic susceptibility~\cite{Wieniak2005}. Furthermore, entanglement in many-body systems of bosons and fermions can be indicated through off-diagonal long-range order (ODLRO), a microscopic property that signals quantum coherence on a macroscopic scale~\cite{Yang1962,Penrose_BEC}, and whose onset underlies phenomena such as exciton condensation~\cite{Safaei2018,Garrod1969,Schouten2022,Schouten2023,Adolphs2006,Mazziotti2012,Schouten2025,Sager2022,PayneTorres2024,Dijkstra2019,Mattioni2021,Giannini2022,Mueller2018,High2012}, superconductivity~\cite{Sager2022_2,Aly_2015}, and superfluidity~\cite{Yang1962,Ma2024}. Real-world quantum materials, however, often involve interacting mixtures of fermions and bosons, where richer many-body behaviors may emerge~\cite{Warren2024,Warren2025,Macridin2018,Lamata2014,Casanova2011,Barry2020,Pang2020,Jiang2009,Udvarhelyi2017,Schouten2025_1}. Such mixed-particle quantum systems—encompassing phonons, photons, and gluons—are pervasive across physics, serving as effective models in condensed-matter and chemical systems~\cite{Holstein1959,Berger1995,Engelsberg1963,Melnikov2001} and forming the natural language of quantum field theory~\cite{Fradkin1994}. Yet existing entanglement witnesses, typically based on spin fluctuations or single-particle observables~\cite{Toth2012,Hyllus2012,Amico2008,Horodecki2009}, fail to capture the collective coherence that characterizes ODLRO, limiting their ability to identify and quantify many-body order in mixed-particle environments.

\begin{figure}[ht!]
    \centering
    \includegraphics[width=\linewidth]{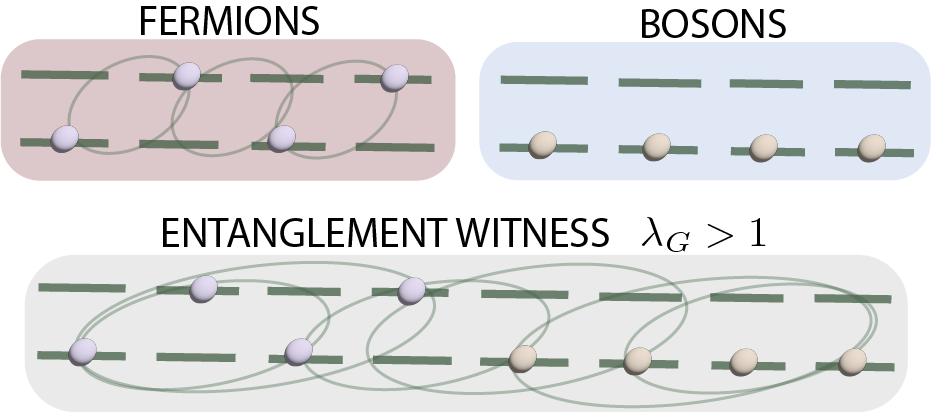}
    \caption{Illustration of entanglement witnessing in a mixed fermion-boson system, where the fermionic degrees of freedom are initially strongly correlated. The largest eigenvalue of the fermion and boson blocks of the mixed-particle particle–hole reduced density matrix, $\lambda_G$, captures entanglement arising in the fermionic and bosonic modes, respectively.}
    \label{fig:witness}
\end{figure}

In this Letter, we present a framework for witnessing entanglement in mixed-particle quantum systems by detecting and quantifying ODLRO. The theory reveals that ODLRO serves as a unifying signature of macroscopic entanglement in systems containing both fermions and bosons, allowing the identification of long-range order within each subsystem as well as across the entire mixed-particle system. The witnesses quantify the respective contributions of the fermionic and bosonic sectors and hence provide a vivid picture of how coherence is distributed between particle types as correlations strengthen. They are defined in terms of the large eigenvalues of the fermion and boson blocks of the mixed-particle particle–hole reduced density matrix. Notably, within the boson sector, a large eigenvalue indicates that entanglement can drive a correlated condensation of particle–hole pairs—distinct from the statistics-driven single-particle condensation of conventional Bose–Einstein condensation. We demonstrate these principles using an expanded Lipkin–Meshkov–Glick spin model, showing how the transition from ODLRO in only one particle type to ODLRO in both particle types captures the onset of collective entanglement in a mixed-particle environment (refer to Fig.~\ref{fig:witness}). This approach establishes a general and scalable route to characterizing entanglement in mixed-particle quantum systems and suggests potential applications in quantum sensing—where entanglement enhances measurement sensitivity—and in quantum communication, where coherence may enable robust information transfer and storage.

{\em Theory---} The emergence of ODLRO can be characterized through the appearance of dominant eigenvalues in the system’s reduced density matrices (RDMs). For bosonic systems, Penrose and Onsager showed that Bose–Einstein condensation corresponds to an eigenvalue greater than one in the one-boson RDM~\cite{Penrose_BEC}. Analogously, Yang~\cite{Yang1962} and Sasaki~\cite{Sasaki1965} demonstrated that a large eigenvalue in the particle–particle RDM indicates condensation of fermion pairs into a single two-particle quantum state, signaling the onset of ODLRO. 

Exciton condensation represents a particle–hole analogue of this phenomenon, where below a critical temperature, particle–hole pairs collectively occupy a single excitonic state, resulting in exciton superfluidity~\cite{keldysh_2017,Fil2018,PayneTorres2025}. Such a phase enables dissipationless energy transfer at elevated temperatures, a long-standing goal in superconductivity research~\cite{EM2004,Butov2003,Kasprzak2006,Min2008,Li2017,Liu2017,Ma2021,Gao2023}. The onset of ODLRO is captured by the particle–hole RDM,
\begin{equation}
    ^2G^{ij}_{kl} = \bra{\psi}(\hat{a}_i^{\dagger}\hat{a}_j - {}^{1\!}D^{i}_{j})^{\dagger}(\hat{a}_l^{\dagger}\hat{a}_k - {}^{1\!}D^{l}_{k})\ket{\psi},
\end{equation}
where $^1D$ is the one-particle RDM, $\hat{a}^{\dagger}_{i}$ and $\hat{a}^{}_{i}$ denote the creation  and the annihilation operators, and $^2G\geq0$ is a positive semidefinite matrix. 

An eigenvalue of the particle–hole RDM exceeding one indicates multiple particle–hole pairs occupying the same excitonic mode, i.e., exciton condensation~\cite{Garrod1969,Safaei2018}. Here, we propose a mixed-particle model capable of exhibiting simultaneous ODLRO signatures in both fermionic and bosonic sectors. By tracing out either the bosonic or fermionic degrees of freedom, the corresponding fermionic or bosonic subblocks of the particle–hole RDM are obtained as
\begin{align}
    ^2_{f}G^{ij}_{kl} &= \bra{\psi}(\hat{f}_i^{\dagger}\hat{f}_j^{} - {}^{1\!}_{f}D^{i}_{j})^{\dagger}(\hat{f}_l^{\dagger}\hat{f}_k^{} - {}^{1\!}_{f}D^{l}_{k})\ket{\psi}, \label{eq:fermionic-G}\\
    ^2_{b}G^{ij}_{kl} &= \bra{\psi}(\hat{b}_i^{\dagger}\hat{b}_j - {}^{1\!}_{b}D^{i}_{j})^{\dagger}(\hat{b}_l^{\dagger}\hat{b}_k^{} - {}^{1\!}_{b}D^{l}_{k})\ket{\psi}, \label{eq:bosonic-G}
\end{align}
where $\hat{f}^{\dagger}_{i}$ and $\hat{f}^{}_{i}$ and $\hat{b}^{\dagger}_{i}$ and $\hat{b}^{}_{i}$ are fermionic and bosonic creation and annihilation operators, respectively.

The largest eigenvalue in each subblock, denoted $\lambda_G$, quantifies the occupancy of the lowest particle–hole mode and, therefore, the extent of ODLRO~\cite{Garrod1969,Safaei2018}. For instance, $\lambda_G=1.5$ indicates the onset of long-range order, while $\lambda_G=2$ corresponds to two correlated pairs. This eigenvalue-based metric has been successfully applied to quantify ODLRO in a variety of quantum and molecular systems~\cite{Liu2025, Sager_2020, Safaei2018, Sager2022, Schouten2021, Schouten2023, Schouten.2023, Sager2022_2}. 

The general upper bound on $\lambda_G$ is given by~\cite{Garrod1969,Schouten2022}
\begin{equation}
    \lambda_G \leq \frac{N(r - N)}{r},
\end{equation}
where $N$ is the number of particles and $r$ is the rank of the single-particle basis. For a mixed system with $N$ fermions and $N$ bosons occupying $2N$ total levels, this bound becomes
\begin{equation}
    \lambda_G \leq \frac{N(2N - N)}{2N} = \frac{N}{2}.
\end{equation}
The maximal large eigenvalue in the particle-hole RDM occurs if the order of the system extends over all particles in the system, hence referred to as long-range order. Because the particle-hole RDM scales linearly with the system size, its largest eigenvalue cannot scale faster than linear in the number of particles. A similar bound has previously been established for the largest eigenvalue of the two-particle cumulant~\cite{Aly_2015,Schouten2022,Liu2025}. 

The composite Hamiltonian describing such a system has the general form $\mathcal{H} = \mathcal{H}_f + \mathcal{H}_b + \mathcal{H}_i,$ where $\mathcal{H}_f$ and $\mathcal{H}_b$ describe the fermionic and bosonic subsystems, and $\mathcal{H}_i$ mediates their interaction. When particle–hole exchange is allowed between these sectors, $\lambda_G$ serves as a unified witness of ODLRO in both fermionic and bosonic degrees of freedom. This framework captures how long-range correlations emerge and redistribute with varying interaction strength, offering a systematic means to characterize entanglement and coherence in correlated mixed-particle environments.

{\em Results---} The Lipkin–Meshkov–Glick (LMG) model is an exactly solvable framework describing spin-$\frac{1}{2}$ particles in a two-level system, where scattering terms facilitate the excitation and de-excitation of particle pairs between the lower and upper levels~\cite{Lipkin_model}. This model has been used extensively to demonstrate fundamental aspects of collective particle behavior~\cite{Schouten2024}, illustrate limitations of mean-field approaches~\cite{Mazziotti1998,M2004,Prez1988,Stein_2000}, and has even been implemented on a quantum computer~\cite{Cervia2021}. The LMG quasispin model for $N$ fermions contains two energy levels, \{-{$\frac{\varepsilon}{2}$ and {$\frac{\varepsilon}{2}$\}, each hosting $N$ states that are energetically degenerate. The second-quantized Hamiltonian is given by
\begin{align}
    \mathcal{H}_f =\ & \frac{\varepsilon}{2} \left[
    \sum_{i=N+1}^{2N} \hat{f}_i^\dagger \hat{f}_i^{}
    - \sum_{i=1}^{N} \hat{f}_i^\dagger \hat{f}_i^{} \right] \nonumber \\
    & + \frac{V}{2} \sum_{p,q=1}^{N} \left( \hat{f}_p^\dagger \hat{f}_q^\dagger\, \hat{f}_{q+N}^{} \hat{f}_{p+N}^{} + \mathrm{h.c.} \right), \label{eq:lmg}
\end{align}
where $V$ is the interaction strength between pairs of scattered particles. In the large-$N$ limit, this model exhibits a normal–to–deformed quantum phase transition---corresponding to a change from a symmetry-preserving to a symmetry-broken collective mean-field state---which, while not the focus of the present work, appears as a smooth crossover for the finite system sizes considered here~\cite{Lipkin_model}.  When $V=0$, the system behaves as fully non-interacting. As $V$ increases, interactions become more significant, ultimately producing a macroscopically large $\lambda_G$ for strongly interacting systems with large particle number $N$. When the correlation term $V$ is sufficiently larger than the energy term, $\varepsilon$, i.e., in the strong correlation limit, maximum signature of ODLRO, $\lambda_G=\frac{N}{2}$ can be obtained~\cite{Garrod1969}.

Since $N$ particles can occupy each level, the LMG model may also be interpreted as a two-level bosonic system. The Hamiltonian for the hard-core boson formulation of the model, $\mathcal{H}_b$, may be obtained by replacing the fermionic operators in Eq.~(\ref{eq:lmg}) with bosonic creation and annihilation operators $\hat{b}_{i}^{\dagger}$ and $\hat{b}_{i}^{}$, respectively. Similar to the fermionic modes, ODLRO signature in the bosonic degrees of freedom is reflected by a large $\lambda_G$ value in strongly interacting systems. 

Here, we propose a model system capable of capturing the development of long-range correlations in fermionic and bosonic degrees of freedom simultaneously. The model keeps the structures of the fermionic and bosonic LMG Hamiltonians, $\mathcal{H}_f$ and  $\mathcal{H}_b$, respectively, with an additional interaction term, tuned via parameter $\mu$, which allows for a co-occurring particle excitation in one particle type and a particle de-excitation in the other. The interaction is described with the following Hamiltonian,
\begin{align}
    \mathcal{H}_i =  \frac{\mu}{2}\sum_{p,q=1}^{N}\bigl(\hat f_{p+N}^\dagger\,\hat f_{p}^{}\,\hat b_{q+2N}^\dagger\,\hat b_{q+3N}^{}+\mathrm{h.c.}\bigr).
    \label{eq:hamiltonian}
\end{align}
Importantly, the model can also be adapted to capture the interaction of fermions with other bosonic collective excitations in solids (e.g., spin, orbital, charge, etc.) or to capture interactions between two types of bosons (or two types of fermions).

\begin{figure}
    \centering
    \includegraphics[width=\linewidth]{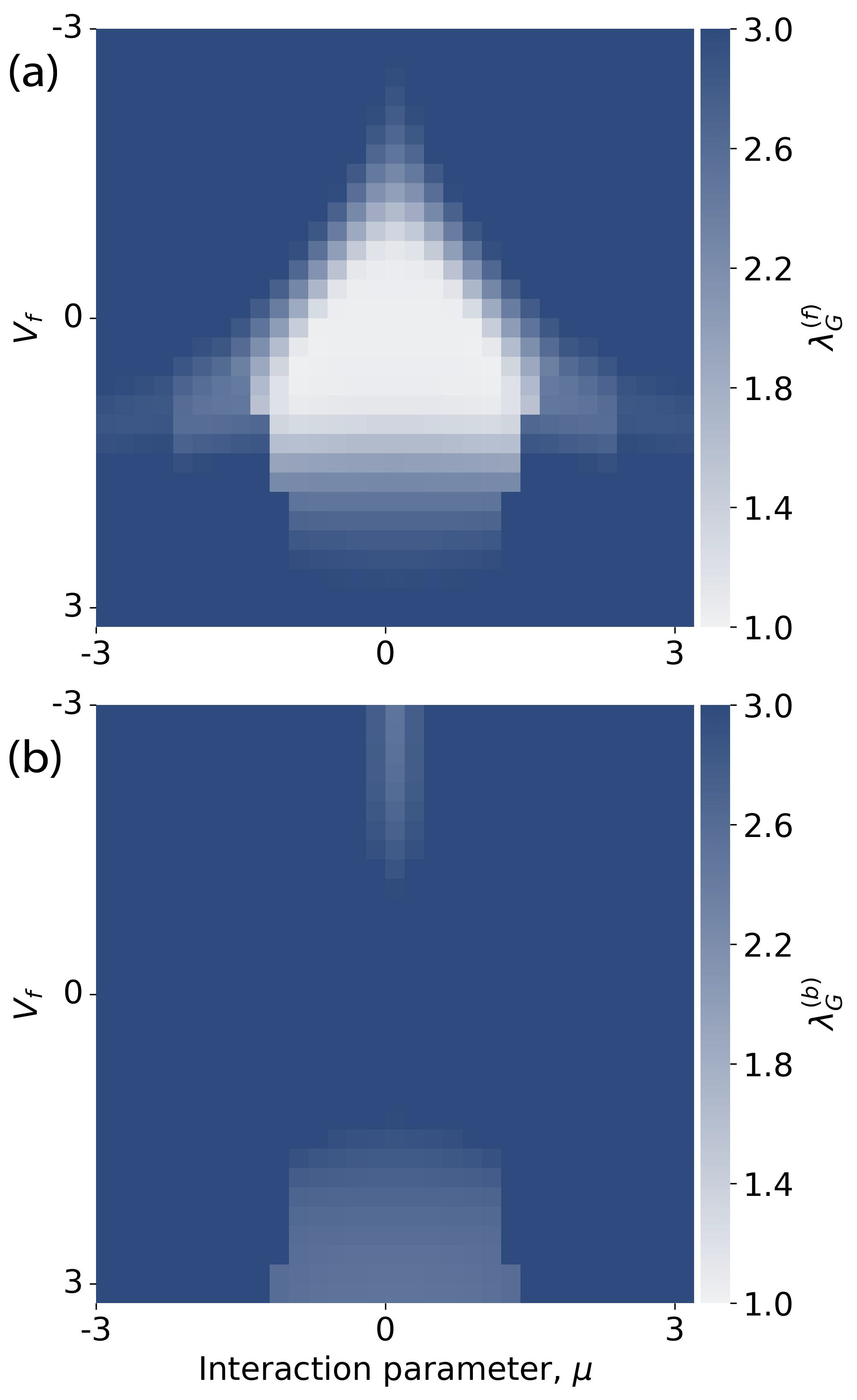}
    \caption{Heatmaps of the entanglement witness (a) $\lambda_G^{(f)}$ and (b) $\lambda_G^{(b)}$ for the fermionic and bosonic particle sectors (f) and (b), respectively, shown as functions of the fermionic correlation parameter $V_f$ and fermion-boson interaction parameter $\mu$, with $V_b = -2$. The system consists of 6 fermions and 6 bosons in 24 orbitals with $\varepsilon_f = \varepsilon_b= 5$. Darker regions indicate a stronger entanglement witness signal, reflecting a greater extent of ODLRO within the corresponding particle degrees of freedom.}
    \label{fig:V1vsmu-high-strength}
\end{figure}

Since the entanglement witness $\lambda_G$ is independent of particle statistics, evaluating it from the modified particle–hole RDM after tracing out one particle type reveals signatures of ODLRO within that sector, while still accounting for its entanglement with the rest of the system. Thus, a parameter set {$\varepsilon_f$, $V_f$, $\varepsilon_b$, $V_b$, $\mu$} directly determines the extent of ODLRO formation in both fermionic and bosonic modes of the mixed-particle quantum system. The interaction parameter $\mu$ allows strong correlations originating in one particle type to interact with correlations in the other, with $\lambda_G$ providing a quantitative measure of the entanglement established across the mixed-particle system.

To illustrate how ODLRO, a special case of entanglement, manifests in a mixed-particle system, we first consider a system of 12 particles (6 fermions and 6 bosons) in 24 orbitals~\footnote{All data and analysis scripts are openly available at the following web address: \url{https://github.com/damazz/Mixed-Particle-Entanglement-Witnesses.git}}. Figure~\ref{fig:V1vsmu-high-strength} shows the entanglement witness $\lambda_G$ for the ground state of the system as a function of the fermionic correlation parameter, $V_f$, and fermion-boson interaction parameter, $\mu$, with all other parameters fixed. Darker blue regions indicate stronger ODLRO. Panel~(a) shows $\lambda_G^{(f)}$ for the fermionic sector. As $V_f$ increases in magnitude, ODLRO strengthens, reaching its maximum at nonzero values of $\mu$. When $\mu=0$, fermions remain weakly correlated, except for large $V_{f}$, and display minimal ODLRO, as indicated by the lighter regions in the plot. Panel~(b) shows $\lambda_G^{(b)}$ for the bosonic sector, which is strongly interacting due to $V_b=-2$, maintaining strong ODLRO across the parameter space. The results establish $\lambda_G$ as an entanglement witness capable of both detecting the presence of long-range correlations within bosonic and fermionic degrees of freedom, and revealing when such coherence extends across the mixed-particle system.

\begin{figure}[ht!]
    \centering
    \includegraphics[width=\linewidth]{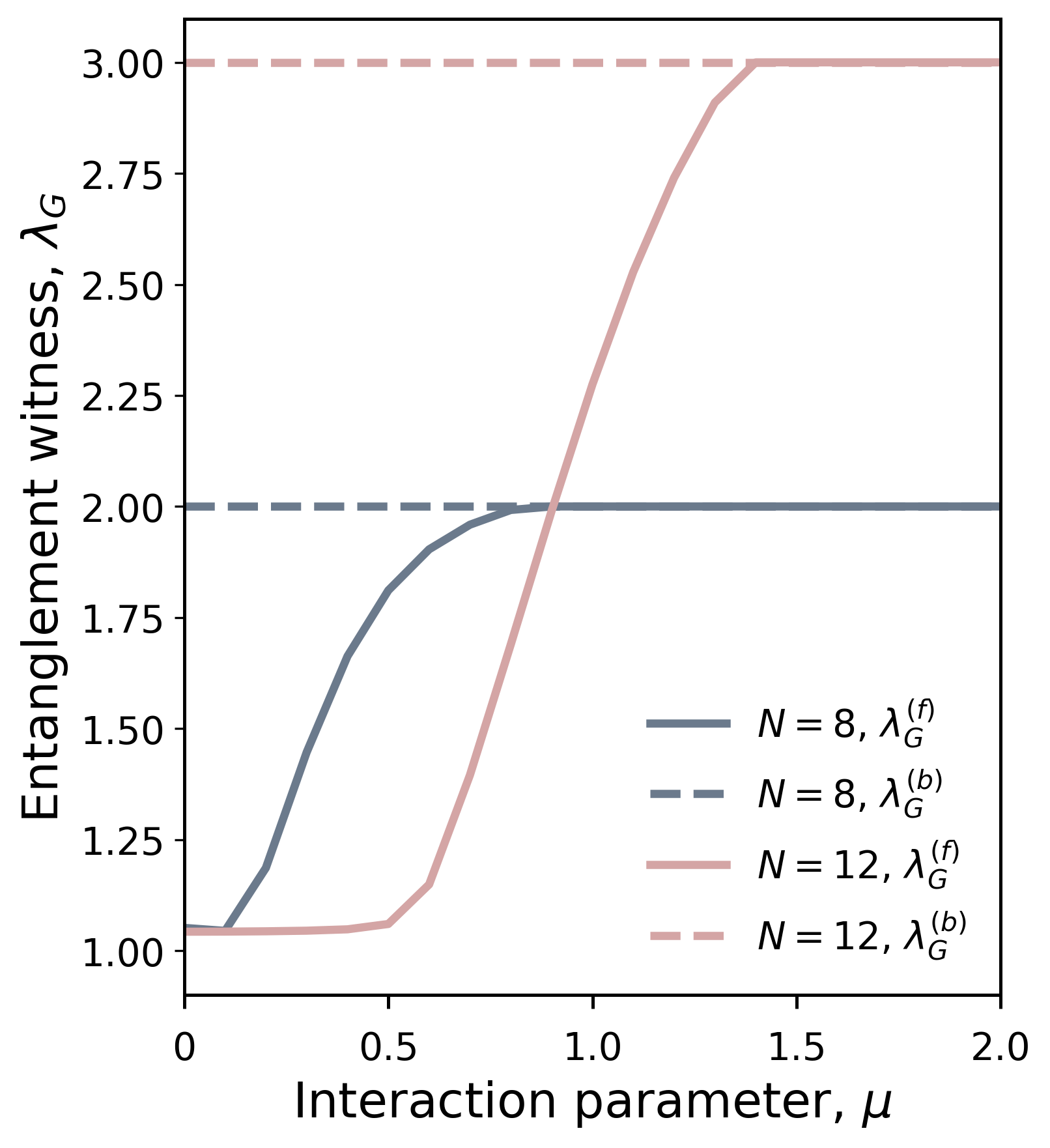}
    \caption{Entanglement witnesses $\lambda_G^{(f)}$ (solid lines) and $\lambda_G^{(b)}$ (dashed lines) for fermions and bosons, respectively, in systems with 8 particles---4 fermions and 4 bosons (blue)---and 12 particles---6 fermions and 6 bosons (pink)---as functions of the interaction parameter $\mu$. For the 8-particle system $V_f = -0.4$, $V_b = -2.0$, and $\varepsilon_f = \varepsilon_b = 1$; for the 12-particle system $V_f = -0.4$, $V_b = -2.0$, and $\varepsilon_f = \varepsilon_b = 5$. In both systems, the fermions, though initially non-interacting, develop increasing ODLRO with stronger interactions $\mu$.}
    \label{fig:V1vsmu-low-strength}
\end{figure}

We next examine a scenario where the fermionic sector (f) is initially non-interacting and exhibits no intrinsic ODLRO ($V_f \ll \varepsilon_f$), while the bosonic sector (b) possesses strong internal interactions ($V_b \gg \varepsilon_b$) sufficient to sustain maximal ODLRO, as indicated by $\lambda_G^{(b)}$. Figure~\ref{fig:V1vsmu-low-strength} presents $\lambda_G$ as a function of the interaction parameter $\mu$ between the two particle sectors. With $V_b \gg \varepsilon_b$, bosonic sector retains its ODLRO independently, reaching maximal $\lambda_G^{(b)}$ values across all $\mu$ (dashed lines in Fig.~\ref{fig:V1vsmu-low-strength}). In contrast, the fermionic sector, though initially lacking correlations, progressively develops ODLRO as $\mu$ increases, as reflected in the rising $\lambda_G^{(f)}$ values (solid lines in Fig.~\ref{fig:V1vsmu-low-strength}). This behavior shows that $\lambda_G$ can effectively witness the onset of collective entanglement in a subsystem, such as the fermionic degrees of freedom, that is fully driven by the interaction Hamiltonian.

Finally, we examine a scenario where the two particle sectors operate on distinct energy and interaction scales that differ by roughly an order of magnitude, as in the case of electrons and phonons, yet are initially either both uncorrelated or both strongly correlated. Figure~\ref{fig:electron-phonon-transfer} shows results for the mixed-particle LMG model containing 12 particles (6 electrons and 6 phonons) in 24 orbitals, with an order-of-magnitude energy difference between electronic and vibronic degrees of freedom ($\varepsilon_e = 3.0$, $\varepsilon_p = 0.3$). Electron–phonon and electron–photon interactions are important for applications in spintronics~\cite{Thomas2016}, catalysis~\cite{CamposGonzalezAngulo2019, KnaCohen2019}, and quantum information processing~\cite{Jahnke2015,Ladd2010}. Dashed lines correspond to both sectors starting uncorrelated ($\lambda_G = 1$), where maximal ODLRO ($\lambda_G = 3$) appears only at strong interaction strength ($\mu > 0.5$). In contrast, the solid lines represent sectors initially strongly correlated ($\lambda_G \approx 2$), reaching the same maximal $\lambda_G$ at a much lower interaction threshold ($\mu > 0.2$). This contrasting behavior suggests that uncorrelated particle-type environments may require stronger interaction to develop coherent order, whereas initially correlated environments appear more responsive to moderate interaction strengths. Moreover, we also observe for both the correlated and uncorrelated initial conditions that the onset of ODLRO in the phonons lags behind the onset in the electrons. Such results imply that the size of the energy gap influences how coherence emerges and distributes between different degrees of freedom in mixed-particle systems. A related enhancement of excitonic coherence has been observed in WSe$_2$/MoS$_2$ moiré heterobilayers, where fermion–boson interactions and the joint modulation of electron filling and exciton density have been linked to tunable interlayer exciton emission~\cite{Tan2025}.

\begin{figure}[ht!]
    \centering
    \includegraphics[width=\linewidth]{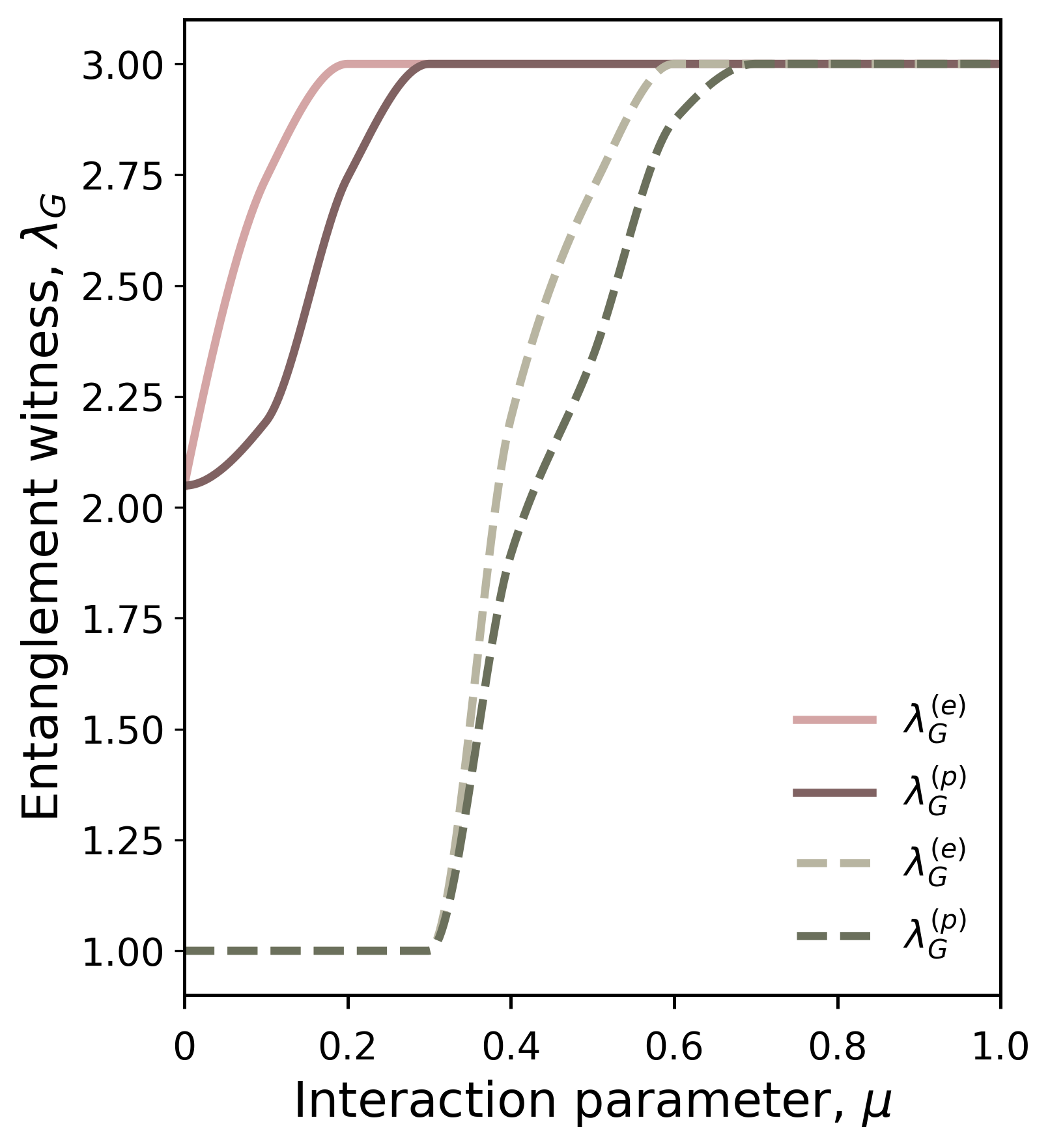}
    \caption{Entanglement witness values $\lambda_G^{(e)}$ and $\lambda_G^{(p)}$ for the electron–phonon model system for 12 particles---6 electrons and 6 phonons---in 24 orbitals as functions of interaction parameter $\mu$. Dashed lines show both particle type environments initially uncorrelated ($V_p = 0.0$, $\varepsilon_p = 0.3$; $V_e = -0.08$, $\varepsilon_e = 3.0$), reaching maximal $\lambda_G = 3$ only in the strong-coupling regime ($\mu > 0.5$). Solid lines show both particle type environments initially strongly correlated ($V_p = -0.08$, $\varepsilon_p = 0.3$; $V_e = -0.8$, $\varepsilon_e = 3.0$), saturating to maximal $\lambda_G$ at modest coupling ($\mu > 0.2$).}
    \label{fig:electron-phonon-transfer}
\end{figure}

{\em Discussion and Conclusions---} Here, we present an entanglement witness for detecting ODLRO in mixed-particle quantum systems. Our results show that the witness successfully quantifies ODLRO---an entanglement-based measure of macroscopic coherence---in systems of up to 12 particles in 24 orbitals, across a broad range of ground-state wave functions generated from diverse parameter sets. In addition to its relevance to exciton condensation in mixed-particle systems, the framework has potential applications in quantum memory, computing, and sensing. In memory, the ability to capture and quantify ODLRO may aid in designing local storage devices where spin–photon entanglement in nanophotonic platforms supports scalable quantum repeater networks~\cite{Bhaskar2020}. In computing, it could inform the development of hybrid fermion–boson architectures such as silicon-based dopant-array analog simulators~\cite{Rad2024}. In sensing, the framework may provide new insight into how fermionic and bosonic degrees of freedom collectively respond to external magnetic or electric fields~\cite{Payne2025}---a direction we aim to explore in future work.

The results further show that the proposed entanglement witness reliably captures how ODLRO distributes between fermionic and bosonic particle types as the interaction strength between them is varied. We find that initially uncorrelated particle types require strong interactions between the fermions and bosons to develop long-range order, whereas those with pre-existing correlations exhibit enhanced ODLRO even under moderate interactions. The relative energy scales of the two particle types are also found to influence this behavior. Together, these findings establish the witness as a robust and quantitative tool for characterizing entanglement in mixed-particle quantum systems and highlight the key role of interparticle interactions and energetics in shaping collective quantum behavior. The framework is directly relevant to correlated fermion–boson systems such as polaritons (electrons and photons) and polarons (electrons and phonons), where the transfer and storage of entanglement among electrons, photons, and phonons underlie emerging quantum technologies in communication, computation, and sensing.

\textit{Acknowledgments---} D.A.M. gratefully acknowledges the Department of Energy, Office of Basic Energy Sciences, Grant DE-SC0026076. I.A. gratefully acknowledges the NSF Graduate Research Fellowship Program under Grant No. 2140001. 

\textit{Data Availability---} All data and analysis scripts are openly available at: \url{https://github.com/damazz/Mixed-Particle-Entanglement-Witnesses.git}

\bibliography{references}

\end{document}